%% file: main.tex
\documentclass[sigconf]{acmart}
\acmConference[EASE 2025]{The 29th International Conference on Evaluation and Assessment in Software Engineering}{17–20 June, 2025}{Istanbul, Turkey}

\AtBeginDocument{%
  \providecommand\BibTeX{{%
    \normalfont B\kern-0.5em{\scshape i\kern-0.25em b}\kern-0.8em\TeX}}}

\setcopyright{acmcopyright}
\copyrightyear{2024}
\acmYear{2024}
\acmDOI{XXXXXXX.XXXXXXX}

\usepackage{amsmath,amsfonts}
\usepackage{algorithmic}
\usepackage{graphicx}
\usepackage{textcomp}
\usepackage{xcolor}
\usepackage{booktabs}  
\usepackage{xcolor}
\usepackage{multirow}
\usepackage{multicol}
\usepackage{colortbl}
\usepackage{comment}
\usepackage{xspace}
\usepackage{pifont}

\usepackage{xcolor, soul}
\colorlet{lightgray}{gray!30}
\sethlcolor{lightgray} 

\newcommand{\nb}[2]{
	\fbox{\bfseries\sffamily\scriptsize#1}
	{\sf\small$\blacktriangleright$\textit{#2}$\blacktriangleleft$}
}

\newcommand\DDR[1]{\textcolor{orange}{\nb{DDR}{#1}}}
\newcommand{\graytext}[1]{\texttt{\hl{\small #1}}}
\newcommand*{\LLa}{\texttt{LLaMA-3-70B}\@\xspace}
\newcommand*{\LLb}{\texttt{LLaMA-3.1-8B}\@\xspace}

\newcommand*{\gf}{\texttt{GPT-4o-mini}\@\xspace}

\newcommand*{\GH}{GitHub\@\xspace}
\newcommand*{\ie}{i.e.,\@\xspace}
\newcommand*{\eg}{e.g.,\@\xspace}
\newcommand*{\etal}{et al.\@\xspace}

\newcommand{\rqone}{\textbf{RQ$_1$}: \emph{How well has existing SOTA research studied RAG in malicious code detection?}}

\newcommand{\rqtwo}{\textbf{RQ$_2$}: \emph{How do RAG-based methods compare with LLMs?}}

\newcommand{\rqthree}{\textbf{RQ$_3$}: \emph{How effective is few-shot learning in the detection of malicious PyPI packages?}}
	
\begin{document} 

\title{MAL-LLM: Evaluating Large Language Models for Detecting Malicious PyPI Packages}
 
\title{LLMs and RAG for Detecting Malicious PyPI Packages: Are We That Far?} 

\title{LLMs and RAG for Detecting Malicious PyPI Packages} 

\title{Detecting Malicious PyPI Packages: A Case Study with LLMs, RAG, and Few-shot Learning} 

\title{Detecting Malicious PyPI Packages using LLMs: A Case Study with RAG and Few-shot Learning}

\title{Detecting Malicious PyPI Packages: A Case Study with LLMs, RAG, and Few-shot Learning}

\title{Detecting Malicious PyPI Packages with LLMs: Does RAG help?}

\title{Detecting Malicious Source Code in PyPI Packages with LLMs: How Useful Is RAG?}

\title{Detecting Malicious Source Code in PyPI Packages with LLMs: How Useful Can RAG be?}

\title{Detecting Malicious Source Code in PyPI Packages with LLMs: Does RAG Come in Handy?}


\author{Motunrayo Ibiyo}
\affiliation{%
	\institution{University of L'Aquila}
	\city{L'Aquila}
	\country{Italy}
}
\email{motunryoosatohanmen.ibiyo@student.univaq.it}

\author{Thinakone Louangdy}
\affiliation{%
	\institution{University of L'Aquila}
	\city{L'Aquila}
	\country{Italy}
}
\email{thinakone.louangdy@student.univaq.it}

\author{Phuong T. Nguyen}


\affiliation{%
	\institution{University of L'Aquila}
	\city{L'Aquila}
	\country{Italy}
}
\email{phuong.nguyen@univaq.it}


\author{Claudio Di Sipio}


\affiliation{%
	\institution{University of L'Aquila}
	\city{L'Aquila}
	\country{Italy}
}
\email{claudio.disipio@univaq.it}

\author{Davide Di Ruscio}
\affiliation{%
	\institution{University of L'Aquila}
	\city{L'Aquila}
	\country{Italy}
}
\email{davide.diruscio@univaq.it}

\begin{abstract}

Malicious software packages in open-source ecosystems, such as \textit{PyPI}, pose growing security risks. Unlike traditional vulnerabilities, these packages are intentionally designed to deceive users, making detection challenging due to evolving attack methods and the lack of structured datasets. In this work, we empirically evaluate the effectiveness of Large Language Models (LLMs), Retrieval-Augmented Generation (RAG), and few-shot learning for detecting malicious source code. We fine-tune LLMs on curated datasets and integrate YARA rules, \GH Security Advisories, and malicious code snippets with the aim of enhancing classification accuracy. We came across a counterintuitive outcome: While RAG is expected to boost up the prediction performance, it fails in the performed evaluation, obtaining a mediocre accuracy. 
In contrast, few-shot learning is more effective as it significantly improves the detection of malicious 
code, achieving 97\% accuracy and 95\% balanced accuracy, outperforming traditional RAG approaches. Thus, future work should expand structured knowledge bases, refine retrieval models, and explore hybrid AI-driven cybersecurity solutions.

\end{abstract}
\maketitle

\input{src/introduction}

\input{src/background}

\input{src/methodology}

\input{src/results}
\input{src/conclusion}

%
%
%
%
%
%

\begin{acks} 
	
	This paper has been partially supported by the MOSAICO project (Management, Orchestration and Supervision of AI-agent COmmunities for reliable AI in software engineering) that has received funding from the European Union under the Horizon Research and Innovation Action (Grant Agreement No. 101189664). The work has been partially supported by the EMELIOT national research project, which has been funded by the MUR under the PRIN 2020 program (Contract 2020W3A5FY). It has been also partially supported by the European Union--NextGenerationEU through the Italian Ministry of University and Research, Projects PRIN 2022 PNRR \emph{``FRINGE: context-aware FaiRness engineerING in complex software systEms''} grant n. P2022553SL. We acknowledge the Italian ``PRIN 2022'' project TRex-SE: \emph{``Trustworthy Recommenders for Software Engineers,''} grant n. 2022LKJWHC. 
\end{acks}

\bibliographystyle{ACM-Reference-Format}
\bibliography{refs}


\end{document}

%% file: src/introduction.tex
\section{Introduction}
\label{intro}

Open-source repositories have revolutionized software development by providing developers with extensive collections of reusable code, fostering innovation and accelerating software deployment \cite{DBLP:journals/ese/RoccoRSNR21}. However, this accessibility also presents security risks, as adversaries can exploit these repositories by injecting malicious code into widely used libraries or creating deceptive packages that compromise user systems \cite{10.1145/3691620.3695492}. Unlike traditional software vulnerabilities caused by inadvertent errors, malicious packages are deliberately engineered to deceive and exploit users, making their detection particularly challenging.

The recognition of malicious code is becoming increasingly critical due to the proliferation of AI-based systems, which are trained on publicly available repositories and that can power complex applications. Beyond the intrinsic difficulty of distinguishing between benign and malicious packages, research on the nature of adversarial software in package ecosystems remains limited. 

Python has been ranked as the top programming language,\footnote{\url{http://bit.ly/420sCPK}} 
and the number of PyPI packages continues to grow on a daily basis. As Python's adoption increases, so does the prevalence of malicious code \cite{10.1145/3643651.3659898}. While package registry managers have implemented various techniques to detect and mitigate malicious code, multiple studies \cite{9724451, zahan2024shifting, 10298430} have shown that malicious packages still persist within registries or on external servers that act as registry mirrors, making them accessible upon request. Furthermore, 
unlike in C/C++ ecosystems, authors of malicious code in the PyPI environment do not frequently reuse code \cite{10.1007/978-3-031-70896-1_3}. This highlights the highly dynamic nature of malicious code in the Python ecosystem, making detection and mitigation even more challenging.

There exist various datasets for malicious code detection. For instance, a curated dataset \cite{OpenSourceDatasetMaliciousSoftwarePackages} contains 4,700 malicious packages from the PyPI and npm ecosystems, or 
the Backstabbers-Knife-Collection \cite{ohm2020backstabber} includes 5,250 malicious packages, of which 2,496 are Python-based. 
Despite their scale, these datasets lack structured indices specifying the behavioural characteristics of each package, limiting their usefulness for 
threat detection. The absence of well-indexed, contextualized data remains a major issue in security research, impeding efforts to 
classify and mitigate threats. 


The proliferation of Large Language Models (LLMs) in recent years 
has allowed for practical applications in different domains, including software engineering \cite{DBLP:journals/software/Ozkaya23b} (SE). LLMs have been applied to solve a wide range of SE tasks, \eg code generation \cite{10.1145/3672456}, code summarization \cite{10.1145/3643916.3644434}, technical debt detection \cite{10449667}, to name a few. However, while demonstrating the potential for 
a plethora of applications, LLMs have their own limitations. In particular, the lack of data for fine tuning and hallucination could potentially jeopardize 
the deployment of LLMs. To this end, Retrieval-Augmented Generation \cite{10.5555/3495724.3496517} (RAG) has been proposed as a practical solution to deal with LLM shortcomings. RAG is used to incorporate an existing content store into the model, allowing it to enhance its knowledge and respond more accurately. So far, RAG has been applied in different domains, demonstrating a promising performance \cite{10.1145/3643795.3648384}. 
Essentially, a question that may pop up at anytime is: \emph{``How useful is RAG in the context of malicious code detection?''}


\vspace{.1cm}
\noindent
$\triangleright$ \textbf{Solution}. In this paper, we introduce a novel 
approach to the detection of malicious PyPI packages, integrating fine-tuned LLMs, RAG, and few-shot learning. 
The ultimate aim is 
to improve the ability of LLMs to differentiate between safe and harmful PyPI packages, with the aim of contributing to 
enhancing the security posture of open-source ecosystems. Our method addresses the current limitations of both static and dynamic analysis techniques by providing a scalable, context-aware malware classification framework. 

\vspace{.1cm}
\noindent
$\triangleright$ \textbf{Evaluation}. Using a dataset of 1,242 malicious 
and 3,752 benign packages, we test the suitability of two prominent knowledge bases, \ie 
YARA Rules and \GH Security Advisories as the sources for feeding RAG. Moreover, we fine-tune open-source LLMs on the curated dataset 
to improve classification accuracy. 

\vspace{.1cm}
\noindent
$\triangleright$ \textbf{Contribution}. A literature review on the state-of-the-art research shows that our work is the first empirical study investigating the usefulness of RAG in the detection of malicious PyPI packages. More importantly, our study reveals that \emph{the experimental results diverge dramatically from the expectations}, \ie incorporating the aforementioned knowledge bases, 
RAG turns out to be ineffective 
in detecting malicious code. 
This triggers the need for further investigation on the selection of suitable data sources for RAG.

\smallskip
\noindent
$\triangleright$ \textbf{Open Science}. We publish a replication package consisting of the dataset and source code to allow for future research~\cite{Artifacts}.

%% file: src/background.tex
\section{Related Work}
\label{background}

\subsection{Malware Detection Techniques}


\emph{Static analysis} involves examining code without executing it by extracting features, \eg 
opcode sequences, control flow graphs, and file metadata. This method allows for the identification of known malicious patterns and anomalies. 
Singh et al. \cite{singh2021survey} provided an extensive review of ML-based malware detection tools that incorporate static analysis methods, and recent systematic reviews of Android malware detection further detail techniques for analyzing 
bytecode and permissions \cite{pan2020systematic}.


\emph{Dynamic analysis} detects malware by monitoring its behaviour during execution within controlled environments (e.g., sandboxes). This technique captures runtime information such as API and system call sequences, network traffic, and file modifications—features that static analysis might miss, mainly when obfuscation is used. 

\emph{Hybrid approaches} combine static and dynamic analysis to leverage the strengths of both methods while mitigating their limitations. In these techniques, static features (e.g., opcode sequences, and control flow graphs) are integrated with dynamic features (e.g., runtime API calls, and system behaviour) to detect malware comprehensively. Singh \etal \cite{singh2021survey} discussed how such integration improves detection accuracy by compensating for the weaknesses of relying solely on static or dynamic analysis.

\textit{Traditional machine learning models} have been widely used for malware detection. These methods rely on manually engineered features derived from static and dynamic analysis, and use classifiers such as support vector machines, decision trees, or ensemble methods. In fact, 
traditional ML models can still offer competitive performance with effective feature extraction \cite{singh2021survey}.

\textit{Emerging approaches leveraging LLMs} 
analyze code and its descriptive metadata. LLM-based methods capture subtle, context-rich patterns that indicate malicious behaviour by fine-tuning domain-specific datasets and using prompt engineering. In a recent study, Fangzhou \etal \cite{wu2024new} discussed how integrating LLMs into cybersecurity frameworks enhances malware detection by providing sophisticated natural language and code analysis.


\subsection{Applications of LLMs and RAG}


A recent study \cite{fu2022linevul} showed that fine-tuning significantly improves a model's ability to capture domain-specific patterns. 
%
Retrieval-Augmented Generation \cite{10.5555/3495724.3496517} (RAG) is a widely used technique for enhancing LLMs 
by incorporating relevant external sources. This knowledge is retrieved through semantic similarity calculations, improving the model’s ability to generate more accurate and contextually informed responses \cite{gao2023retrieval}.  
Within the security domain, RAG has been explored in several studies for detecting vulnerabilities and cyber threats. For instance, Yu \cite{yu2024retrieval} proposed a technique leveraging RAG to detect vulnerabilities in smart contracts. Similarly, Du \etal \cite{du2024vul} constructed a knowledge base from Common Vulnerabilities and Exposures (CVEs) and augmented LLM prompts with this information to enhance vulnerability detection in code.

Despite these promising applications of RAG, according to our investigation, there has been no research applying this technique to detecting malicious code in software packages. This presents an opportunity to explore RAG's potential in improving the security of open-source ecosystems by identifying and mitigating malicious packages more effectively. 

%% file: src/methodology.tex
\section{Proposed Methodology}

This section describes the proposed approach to the detection of malicious packages using a combination of LLMs, RAG, and few-shot learning. 

\subsection{Incorporating Knowledge Bases with RAG} 

\begin{figure}[t!]
	\centering
	\includegraphics[width=\linewidth]{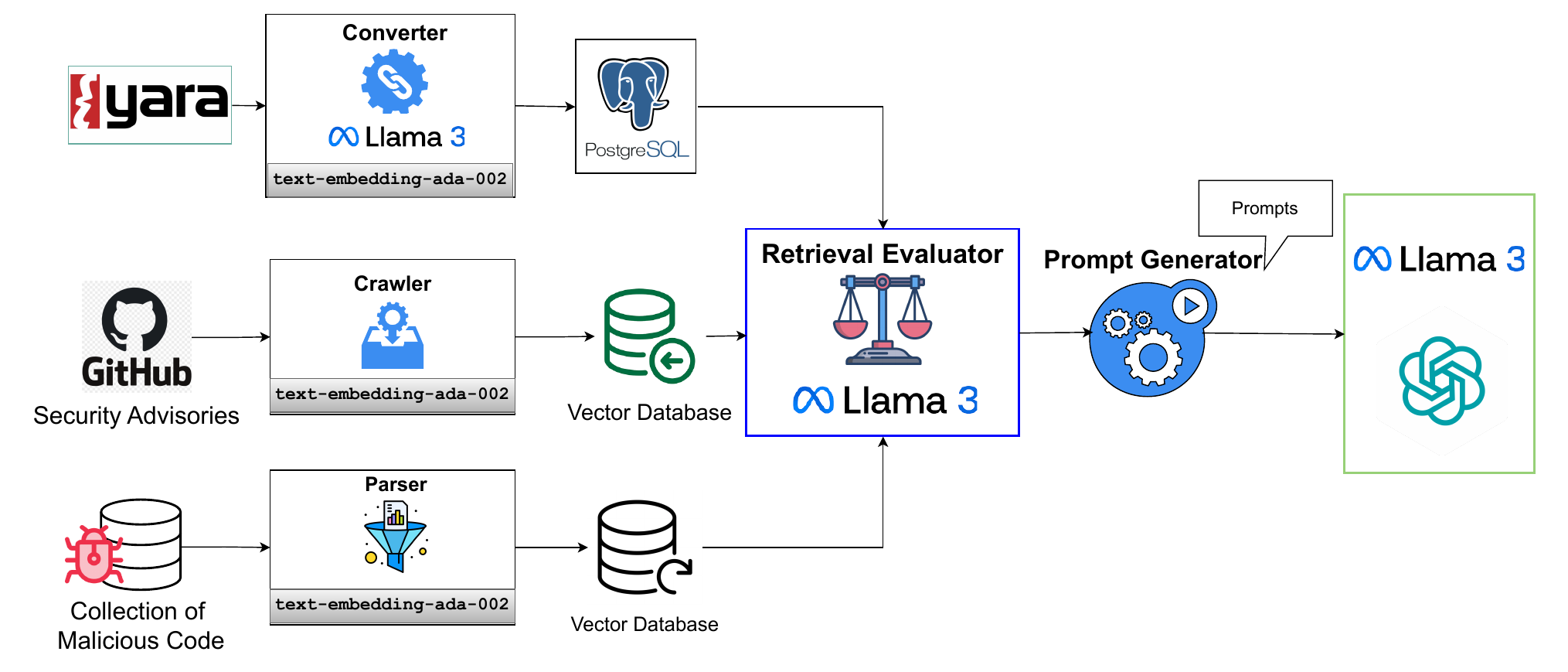}
	\caption{The proposed architecture.}
	\label{fig:Architecture}
\end{figure}

The proposed architecture is depicted in Figure \ref{fig:Architecture}. Given a package, the following 4 key features are considered as input: \emph{(i)} The name of the package; \emph{(ii)} the list of files contained in the package; \emph{(iii)} the content of the \texttt{setup.py} file; \emph{(iv)} the Abstract Syntax Tree (AST) representation of input code. 

The first stage of any RAG system involves constructing the knowledge base (KB) by selecting reliable sources and optimizing retrieval methods \cite{10.5555/3495724.3496517}. 
This task can be challenging due to the lack of standardized documentation regarding malicious packages. Through careful investigation, we identified several sources that are suitable for detecting malicious code:
\emph{(i)} YARA Rules \cite{YARAForge2024}; \emph{(ii)} \GH Security Advisories;\footnote{\url{https://github.com/advisories}} and  \emph{(iii)} Raw code snippets from \texttt{setup.py} contained in the datasets used for experiments. 

\smallskip
\noindent
$\triangleright$ \textbf{YARA Rules Knowledge Base}.
YARA is a widely used tool for identifying and classifying malware through signature-based detection \cite{YARA2024, naik2021embedded}. It enables pattern-matching in files, memory processes, and systems. 
We utilized a curated set of 6,220 rules from the YARA Forge \cite{YARAForge2024}, validated against industry best practices \cite{YARAStyleGuide2024}. 
An LLM is used to transform the rules into human-readable text. Following an existing study~\cite{10298430}, we instruct the LLM to translate the rules into descriptive text to explain what the rules do. Then, the output was structured into documents 
storing metadata such as rule ID, author, and affected OS. Finally, the descriptions were embedded using OpenAI’s \texttt{text-embedding-ada-002} model and stored in a PostgreSQL vector database.

\smallskip
\noindent
$\triangleright$ \textbf{\GH Security Advisories}. 
It is a database for 
vulnerability samples, including CVEs, curated from open-source platforms \cite{GitHubAdvisories2024}. For the PyPI ecosystem, the knowledge base 
contains 3,474 security advisories, but notably, it lacks malware-specific advisories—these are currently available only for the npm ecosystem. The advisories were retrieved via the \GH API, pre-processed and embedded using the \texttt{text-embedding-ada-002} model, and stored in a separate collection within the vector database.

\smallskip
\noindent
$\triangleright$ \textbf{Malicious Code Knowledge Base}. 
To further evaluate the ability of RAG, we embedded and indexed samples of \texttt{setup.py} files from malicious packages. The training set was stored separately to allow for isolated retrieval per knowledge category. Ultimately, we excluded 24 out of 1,158 malicious \texttt{setup.py} files due to their excessive length.

\smallskip
\noindent
\textbf{Retrieval Evaluator}. Corrective Retrieval-Augmented Generation (CRAG) is an improved RAG framework that enhances the retrieval robustness by refined knowledge selection before enhancement \cite{yan2024corrective}.  In our implementation, a retrieval evaluator was introduced to assess document relevance before inclusion in the LLM prompt, and to independently evaluate YARA rules and GitHub advisories. In this way, we are able to append only highly relevant knowledge for the detection. CRAG was tested under two conditions: \emph{(i)} Providing the classifier with the raw code snippet; and \emph{(ii)} Providing only the AST representation of the code (without raw code). CRAG minimizes irrelevant rule inclusion and improves classification accuracy by ensuring only high-quality contextual knowledge is used.

Each knowledge base was tested individually to determine which source provides the most relevant and effective context for LLMs. First, relevant documents are retrieved from the chosen knowledge base. Then these obtained documents are concatenated with the \texttt{setup.py} snippets and prompt template. Finally, the final predictions are generated 
based on the augmented context.

\subsection{Fine tuning with few-shot learning} 


We parsed through 5,887 PyPI software packages, 
using the Python \graytext{ast} library to traverse each package's code and capture basic features, \eg usage of \texttt{subprocess.Popen}, \texttt{os.system} calls with \texttt{bash}, \texttt{eval}, and \texttt{exec}. These extracted features were then transformed into coherent textual descriptions, assigned binary labels (0 for benign, 1 for malicious), and saved into CSV files after splitting the data into training, validation, and test sets. The features obtained from the packages, such as imports and system calls, are mapped to descriptive labels. 
These feature descriptions help improve the model’s ability to detect malicious behaviour based on code operations rather than relying on metadata alone. 

Before training, we loaded and shuffled the CSV files and converted them into a DataDict format compatible with the Hugging Face 
library.\footnote{\url{https://huggingface.co/docs/hub/models-libraries}} Then, we classified the test set to assess the model's baseline performance before fine-tuning. We fine-tuned the pretrained \graytext{\LLb} 
model using these package descriptions. 

For fine-tuning, we initialized the LLM 
model for sequence classification using the 
\texttt{AutoModelForSequenceClassification} library. Since LLaMa is a decoder-only transformer architecture, we replaced its last layer with a linear layer and cast the output to a binary decision (0 for benign and 1 for malicious). The training process involved loading the tokenized text, setting up the model, and training on the training set by minimizing the cross-entropy loss function while monitoring performance on the validation set. We used a learning rate of 1e-4, a batch size of 1, and trained for one epoch. After training, we evaluated the fine-tuned model on the test set to measure its performance. 

%% file: src/results.tex
\section{Evaluation, Results and Discussion}
\label{result}

\subsection{Dataset and Baselines}

For evaluating the approach, we mined data from the PyPI packages in the DataDog malicious software packages dataset \cite{OpenSourceDatasetMaliciousSoftwarePackages}, which contains over 1,929 packages identified as malicious from the Python ecosystem. These packages were processed using Python's Abstract Syntax Tree (AST) to extract relevant code behaviour features, such as system calls, file interactions, and sensitive operations. 
Due to specific issues such as escape code and obfuscation, 
some packages could not be processed, and thus we successfully extracted features from 1,242 packages. 

We used a pre-extracted dataset 
containing 139 malicious and 5,193 benign packages. After applying the same AST-based feature extraction process, we curated features from 93 malicious and 3,752 benign packages \cite{zhang2023malicious}. These features are then converted into textual descriptions that serve as the input for fine-tuning the LLMs.

Example of a textual description is as follows: 
\textit{``start entry selfed\-gamestudy-5.59/setup.py, import process module, import operating system module, use process module call, end of entry.''} 
These descriptions, derived from the features, are used as input for fine-tuning the LLMs, enabling the model to recognize malicious and benign code patterns based on their behaviour.

\begin{figure}[h]
	\centering
	\includegraphics[width=0.80\linewidth]{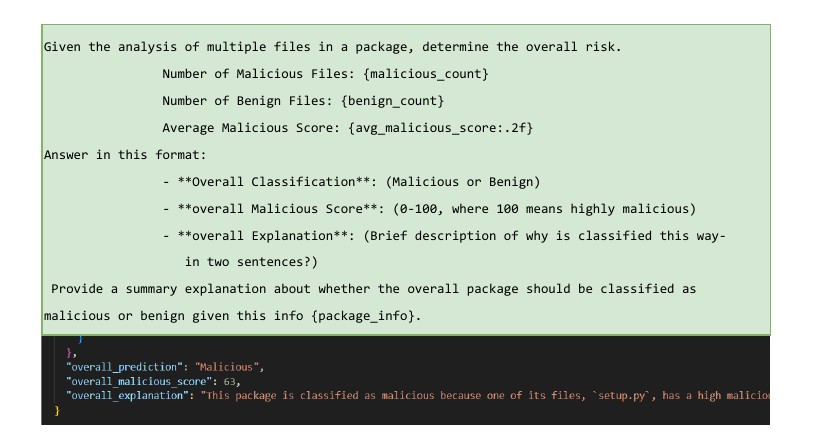}
	\caption{Example of a zero-shot prompt.}
	\label{fig:Prompt}
\end{figure}

The classification process was carried out in two steps. In the first step, each file within a package was classified individually. In the second step, the results from the first step were aggregated to determine the overall classification of the package as **malicious** or **benign**, assigning a malicious score and providing an explanation for the classification.

We used various prompts to query the LLMs, an example of a zero-shot prompt is shown in Figure \ref{fig:Prompt}. Due to space limit, we could not show the few-shot prompts here. Interested readers are kindly referred to the replication package for further details~\cite{Artifacts}.

\subsection{Results}

\vspace{.2cm}
\noindent
$\triangleright$ \rqone~
We performed a four W-questions search \cite{ZHANG2011625} as follows:

\smallskip
\noindent 
\textbf{Which?} We collected relevant peer-reviewed papers from various sources, including conferences and journals in computer science. Our focus was on automated approaches utilizing LLMs and RAG to detect malicious code.

\smallskip
\noindent  
\textbf{Where?} Our literature analysis is centered on computer science papers. To automate this process, we utilized the \textit{Scopus} database\footnote{\url{https://scopus.com}} and leveraged its advanced search and export functions to retrieve pertinent content, including titles, abstracts, and publication venues.

\smallskip
 \noindent  
 \textbf{What?}  For each article, we extracted information from the title and abstract by applying predefined keywords to ensure relevance to our research focus. The list of keywords was defined as follows: \textit{(i)} malicious OR security; \textit{(ii)} "RAG" OR "retrieval-augmented generation"; \textit{(iii)} large AND language AND model OR LLM; \textit{(iv)} classif* OR detect*. It is important to note that all the keywords were combined using the AND operator.
 
 \smallskip
 \noindent  
 \textbf{When?} Given that RAG is a relatively new topic, our search was limited to the past three years, specifically from 2022 to 2025. This time frame allows us to capture the latest developments in the field. Remarkably, we only obtained papers published starting from the beginning of 2024, highlighting the novelty of the topic.

\smallskip
After the initial automatic search, we identified 23 papers that met our criteria. We then reviewed the title and abstract of each paper to ensure their relevance to our research focus: automated detection of malicious code using LLMs with the RAG technique. This process allowed us to select a list of 8 papers. We further refined this list by removing 2 papers that were outside the scope of our study, resulting in a final set of 6 papers, which we reviewed as follows. 

Sheng \etal \cite{sheng_lprotector_2024} proposed LProtector, an automated system for detecting vulnerabilities based on GPT-4o and RAG. First, Common Weakness Enumeration (CWE) rules and corresponding code snippets are embodied in the prompt. RAG leverages the Big-Vul benchmark to enhance the retrieval capabilities of the baseline model. 

Lan \etal \cite{lan_securing_2024} adopted the so-called ``security shift-left'' paradigm, which involves integrating security measures at the outset of the development lifecycle, to augment the robustness of LLM-based systems. In particular, RAG is employed as a secondary defense mechanism equipped with a set of malicious-intent free-text examples relevant to the task. To validate the approach, the authors compare traditional ML models with BERT, showing that RAG is useful in detecting malicious content. 

LLM-CloudSec \cite{cao_llm-cloudsec_2024} is an unsupervised approach that exploits RAG and CWE as an external knowledge base to improve the detection of vulnerabilities for C++ applications. In particular, the system is based on top of two agents, \ie the Detection and the Analysis agents. The former detects possible vulnerabilities in the code using the CWE standards. The latter analyzes the results of the classification to extract detailed descriptions of vulnerabilities. 

Yu \etal \cite{yu_malware_2024} focus on detecting possible security issues in the eBPF Linux Kernel by experimenting with fine-tuning, RAG, and few-shots prompting. First, the system has been monitored using dedicated components that capture raw behavioral log data. Then, the collected logs were grouped using the same process to perform the fine-tuning phase. The RAG module is eventually used to detect possible abnormal behavior using the fine-tuned version of the Qwen model. The conducted ablation study reveals that the introduction of RAG contributes to improving the overall accuracy.

PenHeal \cite{huang_penheal_2024} exploits the counterfactual prompting strategy to autonomously identify and mitigate security vulnerabilities. It consists 
of two modules, \ie Pentest, which detects multiple vulnerabilities within a system, and the Remediation module, which recommends optimal strategies. Moreover, an Instructor module integrates Counterfactual Prompting and RAG to explore multiple potential attack paths. The conducted ablation study reveals that PenHeal outperforms baseline models, PentestGPT and GPT4. 

LLM-Sentry \cite{irtiza_llm-sentry_2024} focuses on handling jailbreaking attacks by combining zero-shot prompting, RAG, and a re-ranker module. In addition, internal knowledge is continually updated using a human-in-the-loop approach that allows the system to detect attacks without modifying the established safeguards. LLM-Sentry shows superior performance compared to three state-of-the-art tools in terms of accuracy and F1-score. 

\vspace{.2cm}
\noindent\fbox{\begin{minipage}{0.98\columnwidth}
		\paragraph{\textbf{Answer to RQ$_1$:}}
		State-of-the-art studies have combined RAG with standard prompting techniques to solve different SE tasks. Nevertheless, none of them used these techniques to detect malicious PyPI packages, posing a remarkable gap in the current literature.        
\end{minipage}}

\vspace{.2cm}
\noindent
$\triangleright$ \rqtwo~

Table~\ref{tab:LLMsAndRAG} shows a comparison of two LLMs, \ie \graytext{\gf} and \graytext{\LLa} in a zero-shot setting with three RAG-based configurations using YARA Rules, \GH Advisories, and Malicious \texttt{setup.py} files. 

\begin{table}[h]
	\centering
	\caption{Comparison of LLMs (zero-shot learning) and \graytext{\LLb} + RAG (YARA Rules), \graytext{\LLb} + RAG (\GH Advisories), \graytext{\LLb} + RAG (Malicious \texttt{setup.py}).} 
	\label{tab:LLMsAndRAG}
	\small 
	\footnotesize
	\begin{tabular}{|l|l|p{1.5cm}|p{1.5cm}|} \hline
		 & \textbf{Model} & \textbf{Benign} & \textbf{Malicious} \\ \hline
		 
		{\multirow{5}{*}{\rotatebox[origin=c]{90}{Precision}}} & \graytext{\gf}   & \textbf{0.92}  &  \textbf{0.98}  \\ \cline{2-4}
		& \graytext{\LLa} & 0.87  &  0.52  \\ \cline{2-4}
		& YARA Rules & 0.83  &  0.48  \\ \cline{2-4}
		& \GH Advisories & 0.83  &  0.39  \\ \cline{2-4}
		& Malicious \texttt{setup.py} & 0.80  &  0.35  \\ \hline

		{\multirow{5}{*}{\rotatebox[origin=c]{90}{Recall}}} & \graytext{\gf}   & \textbf{0.99}  &  \textbf{0.78}  \\ \cline{2-4}
		& \graytext{\LLa} & 0.73  &  0.66  \\ \cline{2-4}
		& YARA Rules & 0.78  &  0.55  \\ \cline{2-4}
		& \GH Advisories & 0.63  &  0.64  \\ \cline{2-4}
		& Malicious \texttt{setup.py} & 0.61  &  0.58  \\ \hline
		
		{\multirow{5}{*}{\rotatebox[origin=c]{90}{F1-score}}} & \graytext{\gf}   & \textbf{0.96}  &  \textbf{0.86}  \\ \cline{2-4}
		& \graytext{\LLa} & 0.79  &  0.59  \\ \cline{2-4}
		& YARA Rules & 0.80  &  0.52  \\ \cline{2-4}
		& \GH Advisories & 0.72  &  0.48  \\ \cline{2-4}
		& Malicious \texttt{setup.py} & 0.69  &  0.44  \\ \hline

	\end{tabular}
\end{table}

The results indicate that \graytext{\gf} consistently outperforms \graytext{\LLb} across all the 
metrics. 
\graytext{\gf} achieves 
high precision (0.92) and recall (0.99) for benign packages, minimizing false negatives. 
Conversely, \graytext{\LLb} exhibits a lower precision (0.52) for malicious packages, suggesting a tendency to produce false positives. The model also struggles with distinguishing between benign and malicious files. 
The results show that \graytext{\LLb} using RAG with YARA rules gets the highest precision (0.83) for benign packages, but struggles with detecting malicious ones due to its reliance on signature-based matching, leading to a lower recall (0.55). In contrast, RAG with \GH Advisories possesses a better recall for malicious samples (0.64), albeit 
a lower precision (0.39), resulting in a higher false positive rate. RAG combined with malicious \texttt{setup.py} yields a lower precision (0.35) but improved recall (0.58). 

These findings suggest that no single knowledge base is sufficient on its own. YARA rules are effective for benign classification, \GH Advisories improve recall for malicious threats, and code-based analysis offers a middle ground. A \textit{hybrid approach} that integrates all three sources may provide a more robust classification system, enhancing both detection accuracy and generalization.

\vspace{.2cm}
\noindent\fbox{\begin{minipage}{0.98\columnwidth}
		\paragraph{\textbf{Answer to RQ$_2$:}} RAG-based methods underperform \graytext{\gf} and \graytext{\LLb} with a zero-shot setting. Among the different RAG configurations, using YARA rules yields the highest accuracy 
		while RAG with \GH Advisories results in the best recall for malicious packages. 
\end{minipage}}





\vspace{.4cm}
\noindent
$\triangleright$ \rqthree~



Before fine-tuning, we evaluated the pre-trained \graytext{\LLb} model on the test set to establish a baseline. Table~\ref{tab:FineTuning} summarizes the classification results before and after fine-tuning using few-shot learning. As seen, before fine tuning, the model’s predictive power was limited, reflected by low precision and recall scores, \eg the precision is lower than 0.72 for both benign and harmful samples. 

\begin{table}[h]
	\centering
	\footnotesize
	\caption{\LLb performance 
		before and after fine-tuning (FT).}
	\label{tab:FineTuning}
	\small 
	\begin{tabular}{|l|l|p{1.5cm}|p{1.5cm}|} \hline
		& \textbf{Model} & \textbf{Benign} & \textbf{Malicious} \\ \hline
		
		{\multirow{2}{*}{Precision}} & Before FT   & 0.66  &  0.72  \\ \cline{2-4}
		& After FT & 0.97  &  0.98  \\ \hline
		
		{\multirow{2}{*}{Recall}} & Before FT   & 0.31  &  0.56  \\ \cline{2-4}
		& After FT & 0.99  &  0.95  \\ \hline
		
		{\multirow{2}{*}{F1-score}} & Before FT   & 0.42  &  0.32  \\ \cline{2-4}
		& After FT & 0.98  &  0.97  \\ \hline
		
	\end{tabular}
\end{table}

It is evident that the model’s classification performance improved substantially after fine-tuning. Table~\ref{tab:FineTuning} shows the metrics on the test set, demonstrating a notable increase in all the evaluation metrics. In particular, the precision scores reach 0.97 and 0.98 for the benign and malicious category, respectively. The same trend is seen with recall, and thus F1-score, \ie they are always greater than 0.95, reaching 0.99 by the benign category. 

\vspace{.2cm}
\noindent\fbox{\begin{minipage}{0.98\columnwidth}
		\paragraph{\textbf{Answer to RQ$_3$:}} Being fine-tuned with few-shot learning, \graytext{\LLb} demonstrates a substantial improvement in the prediction performance compared to that before fine-tuning. 
\end{minipage}}

\subsection{Discussion}

\vspace{.1cm}
\noindent
$\triangleright$ \textbf{Implications}. The experimental results showed that the incorporation of all the three knowledge bases, \ie YARA rules, \GH Advisories, and malicious \texttt{setup.py} files does not help the LLMs improve their prediction performance. This is a counterintuitive outcome as we expected to see a gain in performance when RAG is utilized. This triggers a question about the suitability of the knowledge bases, \ie one may say that YARA Rules and \GH Advisories are not suitable for the detection of harmful packages. However, using the set of malicious \texttt{setup.py} files as the source for RAG also does not contribute to an increase in the accuracy. This implies that the data source itself is not a problem, but the RAG technique is. Future research should address the following issues. First of all, it is necessary to develop a standardized and comprehensive malicious code knowledge base to improve RAG-based solutions. Second, attempts should be made to investigate hybrid models that dynamically combine fine-tuning and retrieval mechanisms to optimize adaptability and efficiency. 
Finally, we should enrich the datasets to include a wider range of malware families and evasion techniques, ensuring more generalizable detection models.

\vspace{.1cm}
\noindent
$\triangleright$ \textbf{Threats to Validity}. Threats to \textit{internal validity} concern two aspects, \ie the literature analysis and the experiment design. Concerning the former, we acknowledge that we may miss relevant works in our review. To mitigate this, we followed well-established guidelines provided \cite{ZHANG2011625} and searched Scopus that contains most of the peer-reviewed venues. Concerning the latter, the selected dataset can lead to biased results as it may not represent all possible malicious threats. In this respect, we leverage CRAG methodology combined with YARA rules to mitigate this issue. \textit{External validity} is related to the generalizability of the whole approach, both in terms of programming languages and examined LLMs. Nevertheless, we considered three different configurations in the experiments by relying on a fine-tuning strategy that can be applied to different LLMs apart from the ones considered. Furthermore, the knowledge base used by CRAG can be customized and applied to additional programming languages.

%% file: src/conclusion.tex
\section{Conclusion}
\label{conclusion}

This study explored the viability of LLMs in detecting malicious software packages, using 
multiple experimental setups, including zero-shot classification, RAG, CRAG, and fine-tuning.
The results indicate that fine-tuning is the most effective approach for detecting malware using LLMs than using RAG, leading to more robust classification results. The experiments highlighted the need for further research in curating credible and structured malicious code knowledge bases that can enhance RAG-based methods. While fine-tuning significantly improves model performance, integrating RAG with optimized knowledge bases such as YARA Rules and GitHub advisories can further refine classification accuracy. Overall, our study demonstrates the effectiveness of fine-tuned and retrieval-augmented LLMs for malicious code detection in the PyPI ecosystem. 